\documentclass[notitlepage,twocolumn,aps,prl,10pt,amsmath,amssymb]{revtex4-1}
\usepackage[T1]{fontenc}
\usepackage{lmodern}
\usepackage[utf8]{inputenc}
\usepackage{bm}
\usepackage{etoolbox}
\usepackage[hidelinks]{hyperref}
\hypersetup{
  pdftitle={Einstein's Equation as a Geometric Encoding of Correlations with Quantum Reference Frames},
  pdfauthor={Eduardo O. Dias}
}

\begin{document}

\title{Einstein's Equation as a Geometric Encoding of Correlations
with Quantum Reference Frames}
\author{Eduardo O. Dias}
\email[]{eduardo.dias@ufpe.br}
\affiliation{Departamento de
F\'{\i}sica, Centro de Ci\^encias Exatas e da Natureza, Universidade Federal de Pernambuco, Recife, Pernambuco
50670-901, Brazil}

\begin{abstract}

In general relativity, the same metric field describes spacetime intervals between events relative to any ideal material reference frame as if the frame were not present. In this sense, the metric encodes these intervals. In quantum mechanics, by contrast, such events are naturally accompanied by correlations with that frame. Motivated by this complementarity, we propose a geometry-information equivalence hypothesis (GIEH), according to which the metric field geometrically encodes the relational information otherwise carried by correlations with a local reference frame (LRF). In a small geodesic ball, the GIEH yields the full nonlinear Einstein equation when ideal-LRF conditions are imposed, without assuming an anti-de Sitter background or an underlying conformal field theory. The framework further allows background correlations to provide an informational origin for the cosmological constant, with their local mutual-information density equal to the Gibbons-Hawking entropy per unit thermodynamic volume.

\end{abstract}

\maketitle

\textit{Introduction}---Diffeomorphism invariance in general relativity (GR) suggests that manifold points lack operational meaning. As emphasized by Einstein, observable spacetime is defined by events---coincidences between the worldlines of material systems~\cite{Einstein2,Giovanelli2021}. More generally, diffeomorphism-invariant observables in gravity can be constructed relative to material reference systems~\cite{rovelli1991,rovelli2004,rovelli2014}. Such systems may be realized as extended material reference frames (ERFs), including scalar fields and dust-like media~\cite{kuchar1991,brown1995,brown1996,whatisRF2024}.

In this context, spacetime intervals between events acquire an operational meaning relative to an ERF. When the ERF contributes negligibly to the stress-energy tensor, we call it ideal; in this regime, its presence does not affect the spacetime geometry~\cite{rovelli1991,whatisRF2024}. More specifically, a local observer of the ERF with four-velocity $u^a$ measures infinitesimal proper-time intervals via $d\tau=-u_a\,dx^a$ along its worldline and spatial distances via $dl=\sqrt{h_{ab}\,dx^a dx^b}$ between neighboring events simultaneous for that observer, where $h_{ab}=g_{ab}+u_au_b$. The displacement $dx^a$ is fixed operationally by worldline coincidences between neighboring ERF observers and other physical systems. Thus, physically equivalent ideal ERFs may have different timelike directions $u^a$ at a given point without changing the spacetime geometry. In this picture, the same $g_{ab}$ describes spacetime intervals relative to \emph{any} ideal ERF as if the frame were not present; in this sense, the metric encodes these intervals.

At the quantum level, by contrast, the material degrees of freedom constituting an ERF must themselves be described quantum mechanically, as in quantum-reference-frame formulations~\cite{giacomini2019,vanrietvelde2020,castro2020,giacominiEQP,giacominiQR}. Accordingly, localization relative to a quantum ERF is naturally accompanied by correlations with its material degrees of freedom~\cite{giacominiQR}. More generally, in relational quantum mechanics~\cite{Rovint}, events establish correlations that make physical properties definite relative to another system. Motivated by this classical-quantum complementarity, we propose here a geometry-information equivalence hypothesis (GIEH), from which the Einstein equation (EE) can emerge.

This work is technically inspired by Jacobson's ingenious thermodynamic and entanglement-based derivations of the EE~\cite{J1,J2}, despite being grounded in a fundamentally different physical principle. In particular, Ref.~\cite{J2} presents the EE as a consequence of the maximal vacuum entanglement hypothesis (MVEH): the vacuum entanglement entropy of a sufficiently small spacelike geodesic ball in a maximally symmetric spacetime is locally maximal. Jacobson assumes that the underlying ultraviolet (UV) physics renders the vacuum entanglement entropy finite~\cite{J2,J3}, with its leading contribution scaling with the boundary area, $S_{\rm UV}\approx\eta A$.

In addition, for a ball radius $\ell$ much smaller than any quantum field theory (QFT) length scale and for first-order perturbations about the vacuum, Jacobson invokes the first law of entanglement, $\delta S=\delta\langle K\rangle$~\cite{J2,Blanco2013}. For nonconformal QFTs, however, $\delta S$ can receive additional small-ball contributions beyond the linear first-law term that dominate as $\ell\to0$~\cite{Casini2016}. Consequently, within this first-order treatment, Jacobson's construction yields the linearized Einstein equation~\cite{Casini2016}, in close analogy with entanglement-based derivations in the anti-de Sitter/conformal field theory (AdS/CFT) correspondence~\cite{EntanglementGravity}. The GIEH, by contrast, requires neither an AdS background nor an underlying CFT and, by retaining nonlinear relative-entropy contributions for small but finite state variations, yields the full nonlinear EE within the small-ball regime.

\textit{The geometry-information equivalence hypothesis}---We now formalize the physical framework underlying the GIEH, setting $c=1$ and using the signature $(-,+,+,+)$. The total system consists of ${\cal S}+{\rm ERF}$, where ${\cal S}$ denotes the physical degrees of freedom other than those constituting the ERF. On a $(d{-}1)$-dimensional spacelike hypersurface $\Sigma$, the total state is described by $\omega$ on a Hilbert space ${\cal H}\simeq{\cal H}^{\cal S}\otimes{\cal H}^{\rm ERF}$, with reduced states $\rho={\rm Tr}_{\rm ERF}\,\omega$ for ${\cal S}$ and $\sigma={\rm Tr}_{\cal S}\,\omega$ for the ERF.

At a point $o\in\Sigma$, let $u^a$ be the unit timelike normal to $\Sigma$, and define a sufficiently small ball $B\subset\Sigma$ of radius $\ell$ centered at $o$, generated by spacelike geodesics from $o$ orthogonal to $u^a$. For $\ell$ much smaller than the local curvature length $L_{\rm curv}$, the geometry within $B$ is locally Minkowskian to leading order; in local inertial coordinates centered at $o$,
$g_{ab}(x)=\eta_{ab}+O(|x|^2/L_{\rm curv}^2)$. With a short-distance QFT cutoff, let ${\cal S}_B$ and the local reference frame (LRF) denote the degrees of freedom of ${\cal S}$ and the ERF supported in $B$, respectively. Their joint state is $\omega_B={\rm Tr}_{\bar B}\,\omega$, where $\bar B$ is the complement of $B$ on $\Sigma$, with reduced states $\rho_B={\rm Tr}_{\rm LRF}\,\omega_B$ for ${\cal S}_B$ and $\sigma_B={\rm Tr}_{{\cal S}_B}\,\omega_B$ for the LRF.

We take LRF states $\sigma_B$ with intrinsic timelike direction at $o$ aligned with the previously defined $u^a$. This may be characterized, for instance, by a local timelike current $\langle J_{\scriptscriptstyle \rm LRF}^a(o)\rangle_{\sigma_B}=j_{\scriptscriptstyle \rm LRF}\,u^a$ or by requiring $u^a$ to be a timelike eigenvector of the LRF stress tensor $\langle T_{ab}^{\scriptscriptstyle \rm LRF}(o)\rangle_{\sigma_B}u^b=-\varepsilon_{\scriptscriptstyle \rm LRF}\,u_a$. For concreteness, one possible realization of a LRF is a spatially smeared particle detector centered at $o$ with four-velocity $u^a$~\cite{Bruno2020}, which can arise as an effective description of a localized QFT probe~\cite{Bruno2024,Torres2024}.

We consider states $\rho$ for which $\rho_B$ differs from the reduced vacuum
state $\rho_B^0$ of ${\cal S}_B$ by a small but finite low-energy perturbation. For example, consider a state $\rho$ describing a particle whose position is spread over scales larger than $B$ but is localized relative to an ERF at the coarse-grained level. The LRF within $B$ detects whether the particle is present by correlating with ${\cal S}_B$. See Supplemental Material~\cite{GIEHsuppl} for a simple toy model. More generally, for a sufficiently small $B$, we assume that an event involving ${\cal S}$ and an ERF is accompanied by correlations between ${\cal S}_B$ and the corresponding LRF. Since we focus on coarse-grained correlations, our discussion is insensitive to the microscopic details of the LRF.

With this setting in mind, recall that, classically, the metric field $g_{ab}$ encodes spacetime intervals between events relative to any ideal LRF since the frame's presence, by definition, does not affect the spacetime geometry. Quantum mechanically, by contrast, an event involving ${\cal S}_B$ and an LRF is accompanied by correlations between them. Taken together, these perspectives suggest that the metric field encodes, in geometric form, the relational information that would otherwise be carried by correlations between ${\cal S}_B$ and an ideal LRF, thereby dispensing with the frame's explicit presence in the geometric description.

Guided by this idea, we formulate the GIEH. Let us define two \emph{a priori} independent entropic quantities. On the one hand, let $S(\rho_B|\sigma_B)=S(\omega_B)-S(\sigma_B)$ denote the conditional entropy~\cite{vedralbook} of ${\cal S}_B$ relative to an LRF, evaluated on a reference background that is locally Minkowskian. On the other hand, let $S_g(\rho_B)$ represent the entropy of ${\cal S}_B$ in the perturbed spacetime geometry $g_{ab}$. We postulate that, in $S_g(\rho_B)$, the metric $g_{ab}$ geometrically encodes the relational information carried by the LRF in $S(\rho_B|\sigma_B)$:
\begin{equation}
S_g(\rho_B)=S(\rho_B|\sigma_B),
\label{GIEH}
\end{equation}
which requires $S(\rho_B|\sigma_B)\geq0$. Since the metric field encodes the spacetime geometry relative to an ideal LRF in $B$, we later show that the GIEH yields the EE in the ideal-LRF limit.

Using $S(\rho_B|\sigma_B)=S(\rho_B)-I(\rho_B:\sigma_B)$, where $I(\rho_B:\sigma_B)$ denotes the mutual information between ${\cal S}_B$ and the LRF, Eq.~(\ref{GIEH}) can equivalently be written as
\begin{equation}
\delta_g S(\rho_B)=-I(\rho_B:\sigma_B),
\label{GIEH2}
\end{equation}
where $\delta_g S(\rho_B)\equiv S_g(\rho_B)-S(\rho_B)$ denotes the entropy difference induced by $\delta g_{ab}$ between the perturbed and reference geometries. Thus, the geometry-induced contribution reduces the entropy of ${\cal S}_B$ by the total LRF correlations geometrically encoded by $g_{ab}$. Equations~(\ref{GIEH}) and (\ref{GIEH2}) are equivalent statements of the central hypothesis of this work. 

In the regulated effective-field-theory setting of Refs.~\cite{J2,Carroll2016}, we separate the local degrees of freedom of both subsystems into ultraviolet (UV) and infrared (IR) sectors, ${\cal H}_B\simeq{\cal H}_B^{\scriptscriptstyle\rm UV}\otimes{\cal H}_B^{\scriptscriptstyle\rm IR}$. The IR sectors contain the ordinary low-energy QFT degrees of freedom on the background geometry, whereas the UV sectors parametrize the short-distance physics. Following Refs.~\cite{J2,Carroll2016}, we assume that, to leading order, the entropy variation induced by $\delta g_{ab}$ is captured by its UV contribution, which is responsible for the leading boundary entanglement: $\delta_g S(\rho_B)=\delta_g S(\rho_B^{\scriptscriptstyle\rm UV})=\eta\,\delta_g A$, where $\rho_B^{\scriptscriptstyle\rm UV}={\rm Tr}_{\rm IR}\,\rho_B$ and $\eta$ is the universal area-entropy density. Substitution into Eq.~(\ref{GIEH2}) then gives
\begin{equation}
\eta\,\delta_g A=-I(\rho_B:\sigma_B).
\label{GIEH3}
\end{equation}
Thus, within the GIEH, the reduction in the area of $\partial B$ is the geometric counterpart of the mutual information between ${\cal S}_B$ and the LRF encoded by $g_{ab}$.

We write $\omega_B=\omega_B^0+\delta\omega_B$, where $\omega_B^0$ is the joint state associated with the local vacuum of ${\cal S}_B+{\rm LRF}$ and $\delta\omega_B$ represents a small but finite IR perturbation. The reduced states of $\omega_B^0$ are $\rho_B^0={\rm Tr}_{\rm LRF}\,\omega_B^0$ and $\sigma_B^0={\rm Tr}_{{\cal S}_B}\,\omega_B^0$. We can then write the mutual information in Eq.~(\ref{GIEH3}) as $I(\rho_B:\sigma_B)=I_B^0+\delta I(\rho_B:\sigma_B)$, where $I_B^0\equiv I(\rho_B^0:\sigma_B^0)$ quantifies the ${\cal S}_B$-LRF correlations already present in the reference vacuum. Thus, using $\delta I(\rho_B:\sigma_B)=\delta S(\rho_B)+\delta S(\sigma_B)-\delta S(\omega_B)$, where $\delta$ denotes the difference with respect to the corresponding reference state, Eq.~(\ref{GIEH3}) becomes
\begin{equation}\label{GIEH4}
\eta\,\delta_g A=-I_B^0-\delta S(\rho_B)-\delta S(\sigma_B)+\delta S(\omega_B).
\end{equation}
Since the perturbation is restricted to the IR sectors, $\delta S(\rho_B)\simeq\delta S(\rho_B^{\scriptscriptstyle\rm IR})$, where $\rho_B^{\scriptscriptstyle\rm IR}={\rm Tr}_{\rm UV}\,\rho_B$.

Rather than truncating at the entanglement first law, we retain the exact relative-entropy identities for the two marginal state variations~\cite{Blanco2013}:
\begin{eqnarray}\label{firstG}
\delta S(\rho_B)&=&\delta\langle K\rangle-S(\rho_B\Vert\rho_B^0),\nonumber\\
\delta S(\sigma_B)&=&\delta\langle K^{\scriptscriptstyle\rm LRF}\rangle-S(\sigma_B\Vert\sigma_B^0).
\end{eqnarray}
Here $K\equiv-\log\rho_B^0$ and $K^{\scriptscriptstyle\rm LRF}\equiv-\log\sigma_B^0$ are the modular Hamiltonians of the marginal reference states. Here $\delta\langle K\rangle$ and $\delta\langle K^{\scriptscriptstyle\rm LRF}\rangle$ are linear in $\delta\rho_B$ and $\delta\sigma_B$, while the relative entropies retain the nonlinear contributions~\cite{Casini2016}.

On the geometric side of Eq.~(\ref{GIEH4}), we take the reference geometry to be a maximally symmetric spacetime (MSS), which is locally Minkowskian to leading order in the small-ball expansion. The fixed-volume variation of the area of $\partial B$ relative to that of a ball with the same volume in this MSS is, to leading order in the small-ball expansion, $\left.\delta_g A\right|_V
=-V_\zeta(G_{ab}+\lambda g_{ab})u^a u^b,$
where $\lambda$ parametrizes the curvature of the MSS through $G^{\rm MSS}_{ab}=-\lambda g^{\rm MSS}_{ab}$~\cite{J2}. At the same order, $V_\zeta\equiv\int_B|\zeta|\,dV
=\frac{\Omega_{d-2}\ell^d}{d^2-1}$
is called the thermodynamic volume associated with the causal diamond whose maximal slice is $B$, and $\zeta$ is its conformal Killing vector normalized to unit surface gravity, $\kappa=1$, as in Refs.~\cite{J2,JacobsonVisser2019}. 

Importantly, the reference MSS and the quantum reference state $\omega_B^0$ play distinct roles: in Eq.~(\ref{GIEH4}), $I(\rho_B:\sigma_B)$ is evaluated on the MSS background, so the correlations present in $\omega_B^0$, quantified by $I_B^0$, already deform the physical spacetime geometry away from the reference MSS.

Substituting Eq.~(\ref{firstG}) and the small-ball expression for $\delta_g A$ into Eq.~(\ref{GIEH4}), we obtain
\begin{eqnarray}\label{EinsteinG}
\eta V_\zeta \left(G_{ab}+\lambda g_{ab}\right)u^a u^b
=I_B^0+\delta\langle K\rangle+\delta\langle K^{\scriptscriptstyle\rm LRF}\rangle\nonumber\\
-S(\rho_B\Vert\rho_B^0)-S(\sigma_B\Vert\sigma_B^0)
-\delta S(\omega_B).
\end{eqnarray}
Equation~(\ref{EinsteinG}) is the general $u^a$-projected GIEH relation before imposing the ideal-LRF conditions, and hence it is not yet the covariant tensorial EE.

Despite their distinct physical foundations, Jacobson's MVEH is recovered as a special limit of the GIEH by first imposing $I(\rho_B:\sigma_B)=\delta S(\rho_B)$ in Eq.~(\ref{GIEH2}), which gives $\delta_{g,\rho}S(\rho_B)=0$, and then restricting to linear order in $\delta\rho_B$~\cite{Casini2016}. Since these additional restrictions have no independent motivation within the GIEH, we do not impose them.

We now show that imposing the ideal-LRF conditions promotes Eq.~(\ref{EinsteinG}) to the covariant tensor form of the EE. To this end, for any state $\rho_B$ of ${\cal S}_B$, we seek joint extensions $\omega_B$ satisfying ${\rm Tr}_{\rm LRF}\,\omega_B=\rho_B$ together with the ideal-LRF conditions specified below.

\textit{The Einstein equation in the ideal LRF limit}---We characterize an ideal LRF by three physically motivated requirements: sufficiently weak coupling to ${\cal S}_B$, negligible stress energy, and no preferred ideal reference frame in the GIEH. First, the coupling to ${\cal S}_B$ must be weak enough that $\rho_B^0$ and $\sigma_B^0$ approximate their isolated reduced vacua in the small-ball regime. Accordingly, the correlations in $\omega_B^0$ must be weak or vanish ($I_B^0\ll1$ or $0$) so that $\omega_B^0\simeq\rho_B^0\otimes\sigma_B^0$, with equality if uncorrelated.

We may therefore evaluate $\delta S(\rho_B)$ and $\delta S(\sigma_B)$, to leading order, about their isolated reduced vacua. For underlying QFT descriptions of ${\cal S}$ and the ERF with UV fixed points, so that they are asymptotically conformal at short distances, and with $\ell$ much smaller than any length scale in either QFT, we consider that the leading small-ball modular contributions take the form~\cite{J2,Casini2016,Speranza2016}
\begin{eqnarray}\label{modular}
\delta\langle K\rangle&=&\frac{2\pi}{\hbar}V_\zeta\left(\delta\langle T_{ab}\rangle u^a u^b-\delta\langle X\rangle\right),\nonumber\\
\delta\langle K^{\scriptscriptstyle \rm LRF}\rangle
&=&\frac{2\pi}{\hbar}V_\zeta\left(\delta\langle T_{ab}^{\scriptscriptstyle \rm LRF}\rangle u^a u^b-\delta\langle X^{\scriptscriptstyle \rm LRF}\rangle\right),
\end{eqnarray}
with all local expectation values evaluated at $o$. These stress-tensor projections are the local energy-density changes of ${\cal S}$ and the LRF relative to their reference vacua.  In contrast to the energy-density projections, $\delta\langle X\rangle$ and $\delta\langle X^{\scriptscriptstyle\rm LRF}\rangle$ are scalars independent of $u^a$.

As exemplified in Refs.~\cite{Casini2016,Speranza2016}, the leading nonlinear contributions associated with the relative-entropy terms in Eq.~(\ref{EinsteinG}) are likewise independent of $u^a$. Specifically, for a CFT deformed by a relevant scalar operator $Y_\Delta$ of scaling dimension $\Delta$, the leading nonlinear contribution has the form $\propto\ell^{2\Delta}\big(\delta\langle Y_\Delta\rangle\big)^2$~\cite{Casini2016}, which is independent of $u^a$. For $\Delta<d/2$, such a contribution can compete with or dominate the usual $O(\ell^d)$ modular contribution in the small-ball limit. In addition, since the reference vacuum state $\omega_B^0$ is homogeneous and does not select a preferred timelike direction, we take $I_B^0$, at fixed $\ell$, to be independent of $u^a$ and constant over spacetime. 

Under these conditions, substituting Eq.~(\ref{modular}) into the GIEH relation in Eq.~(\ref{EinsteinG}) yields
\begin{eqnarray}\label{EinsteinG2}
\eta V_\zeta\left(G_{ab}+\lambda g_{ab}\right)u^a u^b
=
&&\frac{2\pi}{\hbar}V_\zeta
\left(\delta\langle T_{ab}\rangle+\delta\langle T^{\scriptscriptstyle \rm LRF}_{ab}\rangle \right)u^a u^b \nonumber\\
&&+I_B^0-\delta F ,
\end{eqnarray}
where
\begin{align}\label{deltaF}
\delta F
\equiv\,
& \frac{2\pi}{\hbar}V_\zeta \left(\delta\langle X\rangle
+  \delta\langle X^{\scriptscriptstyle \rm LRF}\rangle \right)
\nonumber\\
&+S(\rho_B\Vert\rho_B^0) + S(\sigma_B\Vert\sigma_B^0)
+\delta S(\omega_B).
\end{align}
Note that Eq.~(\ref{EinsteinG2}) places the LRF stress-energy tensor on the same footing as that of ${\cal S}$. The second ideal-LRF requirement is therefore that the LRF stress-energy perturbation be negligible, $\delta\langle T^{\scriptscriptstyle\rm LRF}_{ab}\rangle\simeq0$, so that the LRF does not appear as an independent matter source in Eq.~(\ref{EinsteinG2}).

The final ideal-LRF condition is that the GIEH should not privilege any
particular ideal LRF. Consider a family of ideal LRFs that are physically
equivalent but associated with different timelike directions $u^a$. In this
ideal-LRF limit, any dependence on $u^a$ in Eq.~(\ref{EinsteinG2}),
reflected in the geometry via $\delta_g A$ on the left-hand side, should
originate solely from ${\cal S}$. Any additional $u^a$ dependence from the
LRF sector that survives on the right-hand side would make $\delta_g A$
sensitive to intrinsic properties of the LRF beyond its role in selecting
$u^a$ and the associated ball $B$. The quantity
$\delta\langle T_{ab}\rangle u^a u^b$ satisfies this condition since its
$u^a$ dependence is fixed entirely by the stress tensor of ${\cal S}$ in
$B$. However, $\delta S(\omega_B)$ also depends on the ${\cal S}_B$--LRF
correlations and may introduce precisely such additional $u^a$ dependence,
not fixed by $\rho_B$ alone. Thus, we implement the no-preferred-frame
condition by requiring $\delta S(\omega_B)$ to be independent of $u^a$
for physically equivalent ideal LRFs, which also makes $\delta F$
independent of $u^a$. In Eq.~(\ref{EinsteinG2}), it is convenient to write
$I_B^0-\delta F=(\delta F-I_B^0)g_{ab}u^a u^b$, where
$g_{ab}u^a u^b=-1$.

As a simple sufficient realization of the condition on $\delta S(\omega_B)$,
consider the class of joint states
$\omega_B[\rho_B]=U_B[\rho_B]\,\omega_B^0\,U_B^\dagger[\rho_B]$
for which ${\rm Tr}_{\rm LRF}\,\omega_B[\rho_B]=\rho_B$. Since $\omega_B$
describes an open subsystem, the nonunique unitary $U_B[\rho_B]$ is
introduced solely as an auxiliary parametrization of this class, not as
its physical dynamics. Unitarity then gives $\delta S(\omega_B)=0$, which
is independent of $u^a$.

More generally, note that the ideal-LRF conditions can be formulated directly by
requiring
$\delta I(\rho_B:\sigma_B)
- \frac{2\pi}{\hbar}V_\zeta\,\delta\langle T_{ab}\rangle u^a u^b$
to be independent of $u^a$. Indeed, the right-hand side of
Eq.~(\ref{EinsteinG}) is $I_B^0+\delta I(\rho_B:\sigma_B)$, where
$\delta I(\rho_B:\sigma_B)=\delta S(\rho_B)-\delta S(\rho_B|\sigma_B)$.
Thus, when the stress-tensor term above accounts for all the
$u^a$ dependence of $\delta S(\rho_B)$, the separate $u^a$ independence
of $\delta S(\sigma_B)$ and $\delta S(\omega_B)$ can be relaxed,
provided that any such dependence cancels in their difference,
$\delta S(\rho_B|\sigma_B)=\delta S(\omega_B)-\delta S(\sigma_B)$.

Requiring Eq.~(\ref{EinsteinG2}) to hold at every point $o$ and for
every unit timelike $u^a$ then yields the covariant tensor equation
\begin{align}\label{EinsteinG3}
\eta\left(G_{ab}+\lambda g_{ab}\right)
&=\frac{2\pi}{\hbar}\,\delta\langle T_{ab}\rangle
+\frac{1}{V_\zeta}(\delta F-I_B^0)\,g_{ab}.
\end{align}
At fixed $\ell$, the Bianchi identity, stress-energy conservation, and constant $I_B^0$ give
\begin{equation}\label{lambda}
\lambda=\lambda_0+\frac{\delta F}{\eta V_\zeta},
\end{equation}
where $\lambda_0$ is an integration constant. Thus, $\delta F$ fixes the local reference-MSS curvature scale.

Finally, substituting Eq.~(\ref{lambda}) into Eq.~(\ref{EinsteinG3}) and identifying $\eta=1/(4G\hbar)$, we obtain
\begin{equation}\label{EinsteinGIEH}
G_{ab}+\Lambda g_{ab}
=
8\pi G\,\delta\langle T_{ab}\rangle,
\end{equation}
where
\begin{equation}\label{lambdaGIEH}
\Lambda=\lambda_0+4\hbar G\,\frac{I_B^0}{V_\zeta}
\end{equation}
is the cosmological constant, with contributions from both the integration constant and the background correlations. Because Eq.~(\ref{firstG}) retains the nonlinear relative-entropy contributions of small but finite state perturbations, Eq.~(\ref{EinsteinGIEH}) represents the full nonlinear EE. Note that \(\delta F\) drops out of it.

\textit{Physical implications of the GIEH}---Since the GIEH does not fix $I_B^0$, it leaves open whether the reference joint state $\omega_B^0$ contains background correlations. If not, $I_B^0=0$, and Eq.~(\ref{lambdaGIEH}) reduces to $\Lambda=\lambda_0$, leaving the cosmological constant as an undetermined integration constant as in Ref.~\cite{J2}. A particularly intriguing alternative is that $\Lambda$ is supplied entirely by background correlations locally encoded in the metric, corresponding to $\lambda_0=0$ and $I_B^0\neq0$. Since the ideal-LRF regime requires $I_B^0$ to be small, Eq.~(\ref{lambdaGIEH}) associates $\Lambda$ with the small mutual information present in the local vacuum and encoded by $g_{ab}$ rather than directly with vacuum energy, whose large QFT contributions underlie the cosmological-constant problem~\cite{Weinberg1989}. For $\Lambda$ to be independent of the auxiliary ball size $\ell$, $I_B^0(\ell)$ must scale as $\ell^d$, with the proportionality coefficient to be determined by an explicit microscopic model.

Remarkably, for $\lambda_0=0$, the local mutual-information density $I_B^0/V_\zeta$ fixed by the GIEH through Eq.~(\ref{lambdaGIEH}) admits a natural counterpart at the de Sitter horizon scale. To establish this connection, we first use the fact that the small-diamond and horizon-scale regimes are geometrically connected in the framework of Ref.~\cite{JacobsonVisser2019}: for small de Sitter diamonds, the thermodynamic volume reduces, to leading order, to the $V_\zeta$ used above. In the opposite limit, as the diamond area radius approaches the de Sitter radius $L_\Lambda$, the conformal Killing vector $\zeta$ reduces to the static-patch time-translation Killing field $\xi^{\rm dS}=L_\Lambda\partial_t$, again with $\kappa=1$. On the maximal slice $\Sigma_H$, the corresponding thermodynamic volume is $V_\xi^{\rm dS}=\int_{\Sigma_H}|\xi^{\rm dS}|\,dV=4\pi L_\Lambda^4/3$~\cite{JacobsonVisser2019}.

On the other hand, in four spacetime dimensions, the de Sitter radius satisfies $L_\Lambda^2=3/\Lambda$, and the Gibbons-Hawking entropy of the de Sitter cosmological horizon, expressed in units of $k_B$, is $S_{\rm dS}=3\pi/(\Lambda\ell_{\rm P}^2)$, where $\ell_{\rm P}^2=\hbar G$~\cite{GibbonsHawking1977}. This expression follows from a semiclassical construction entirely independent of the GIEH. Nevertheless, from Eq.~(\ref{lambdaGIEH}), the corresponding local and horizon-scale densities satisfy
\begin{equation}
\label{dSdensityGIEH}
\frac{S_{\rm dS}}{V_\xi^{\rm dS}}
=
\frac{I_B^0}{V_\zeta}
=
\frac{\Lambda}{4\ell_{\rm P}^2}.
\end{equation}
This equation shows that, per unit thermodynamic volume, the local background mutual information coincides exactly with the Gibbons-Hawking entropy. This suggests interpreting the de Sitter entropy as the horizon-scale manifestation of the same background correlation density encoded locally in the metric field. 

Another physical implication of the GIEH concerns $\delta F$ in Eq.~(\ref{lambda}). When $\delta F\neq0$, it is absorbed into $\lambda$ via Eq.~(\ref{lambda}) so that a state perturbation changes the reference MSS in addition to the physical spacetime geometry governed by Eq.~(\ref{EinsteinGIEH}). Nevertheless, since the ideal-LRF conditions do not uniquely determine $\delta F$, the GIEH leaves room for imposing an additional constraint on the correlations encoded by $g_{ab}$, thereby fixing $\delta F$ without modifying the resulting EE. A particularly consequential physical choice is to require the reference MSS to remain fixed under any state perturbation, i.e., $\delta F=0$. In this case, $I(\rho_B:\sigma_B)$, encoded by $g_{ab}$ in Eq.~(\ref{GIEH3}), deforms only the physical geometry. If, in addition, $\lambda_0=0$, the fixed reference spacetime in Eq.~(\ref{lambda}) is Minkowski.

From Eq.~(\ref{EinsteinG3}),
$\delta I(\rho_B:\sigma_B)
=
\frac{2\pi}{\hbar}V_\zeta\,\delta\langle T_{ab}\rangle u^a u^b-\delta F$.
Thus, defining the energy above the vacuum in $B$ by $\delta E=V\,\delta\langle T_{ab}\rangle u^a u^b$, where $V=\Omega_{d-2}\ell^{d-1}/(d-1)$ is the volume of $B$, the condition $\delta F=0$ becomes equivalent to
\begin{equation}
\delta I(\rho_B:\sigma_B)
=
\beta\,\delta E,
\label{constraint}
\end{equation}
where $\beta=2\pi\ell/[\hbar(d+1)]$. Here $\beta$ is fixed by the small-ball modular geometry rather than introduced as a thermodynamic parameter: since $2\pi V_\zeta/\hbar=\beta V$, the stress-tensor contribution to $\delta\langle K\rangle$ is precisely $\beta\delta E$, with the factor $1/(d+1)$ arising from the spatial average of $|\zeta|$ over $B$.

As noted above, if $\lambda_0=0$ and $\delta F=0$ [or Eq.~(\ref{constraint})], the reference spacetime defined in Eq.~(\ref{lambda}) is Minkowski, while Eq.~(\ref{lambdaGIEH}) makes $\Lambda$ entirely determined by $I_B^0$. The total mutual information $I(\rho_B:\sigma_B)$ in Eq.~(\ref{GIEH3}) then deforms the geometry away from Minkowski. Conversely, if $I_B^0=0$ and $\delta F=0$, then $\Lambda=\lambda_0$ in Eq.~(\ref{lambdaGIEH}), so the reference MSS in Eq.~(\ref{lambda}) has the cosmological-constant curvature (de Sitter for $\lambda_0>0$), with $I(\rho_B:\sigma_B)$ deforming the geometry away from it.

Equation~(\ref{constraint}) is reminiscent of energy-correlation tradeoffs in quantum thermodynamics, including bounds of the form $I\leq\beta\Delta E_{\rm total}$ for initially uncorrelated thermal systems~\cite{Bruschi2015}. Supplemental Material~\cite{GIEHsuppl} derives, independently of Eq.~(\ref{constraint}), a QFT modular-correlation inequality that recovers this thermal bound as a particular limit and shows that Eq.~(\ref{constraint}) is compatible with, but generally does not saturate, it. Rather, Eq.~(\ref{constraint}) is the physical requirement that the reference MSS remain fixed, while $\beta$ is independently determined by $\delta\langle K\rangle$.

Finally, the GIEH opens a further possibility in which $g_{ab}$ also encodes a specific UV structure of the ${\cal S}_B$-LRF correlations. As discussed in Supplemental Material~\cite{GIEHsuppl}, for any short-distance regulator $\epsilon >0$, $S^{(\epsilon)}(\rho_B)$ and $I^{(\epsilon)}(\rho_B:\sigma_B)$ are finite but may diverge separately as $\epsilon\to0$. Their UV-divergent parts may nevertheless cancel in
$S^{(\epsilon)}(\rho_B|\sigma_B)
=
S^{(\epsilon)}(\rho_B)
-
I^{(\epsilon)}(\rho_B:\sigma_B)$
so that the $\epsilon\to0$ limit, not considered in the previous formulation, is finite and defines the conditional entropy entering Eq.~(\ref{GIEH}). If, in addition, the resulting finite UV residue takes the area form $\eta A$, the relation $\delta_g S(\rho_B)=\eta\,\delta_g A$ need not be introduced as an independent input associated with otherwise unspecified UV physics, as in Refs.~\cite{J2,Carroll2016}. Instead, it may arise from the UV ${\cal S}_B$-LRF correlation structure together with its geometric encoding by the GIEH.

\textit{Conclusion}---We have proposed the GIEH as an operational and relational bridge between the complementary treatments of events and reference frames in GR and quantum mechanics. Under the ideal-LRF conditions, it yields the full nonlinear EE in the small-ball regime, without requiring an AdS background or an underlying CFT. More broadly, the GIEH permits correlations with ideal LRFs, geometrically encoded by the metric field, to underlie poorly understood features of spacetime geometry, including an informational origin of the cosmological constant and its connection with de Sitter entropy. Developing an explicit microscopic realization is an important next step toward testing these possibilities. Motivated by quantum stress-energy fluctuations, future work could explore an averaged formulation of the GIEH, $\eta\langle\delta_g A\rangle=-I(\rho_B:\sigma_B)$, or, independently, a formulation of Eq.~(\ref{GIEH}) conditioned on ERF records, associating an effective classical geometry $g_{ab}^{(r)}$ with each joint realization $r$ of those records.

\section*{Acknowledgments} \label{sec:acknow}
I am grateful to Bruno de S. L. Torres, Aditya Iyer, and Pedro M. de Moraes for valuable discussions that helped further develop important aspects of this work and for their critical reading of the manuscript.

% Supplemental Material is part of this document, after the main references.
\clearpage
\onecolumngrid
\pdfbookmark[0]{Supplemental Material}{supplemental}
\begin{center}
\large\bfseries Supplemental Material
\end{center}
\phantomsection\label{sec:supplemental}
\setcounter{section}{0}
\setcounter{secnumdepth}{2}
\renewcommand{\thesection}{\Roman{section}}
\setcounter{equation}{0}
\renewcommand{\theequation}{S\arabic{equation}}
% Distinct hyperref destinations when the main-text counters are reset.
\ifdefined\theHequation
  \renewcommand{\theHequation}{supp.\arabic{equation}}
\fi
\ifdefined\theHsection
  \renewcommand{\theHsection}{supp.\Roman{section}}
\fi

This Supplemental Material presents a toy model of localization-induced correlations with a local reference frame, derives modular-correlation inequalities, and discusses a possible relational UV regularization of the conditional entropy. References to Eqs.~(\ref{GIEH}), (\ref{GIEH3}), (\ref{firstG}), and (\ref{constraint}) refer to the main text above; supplemental equations are numbered (S1), (S2), and so on.

\section{ A toy model linking localization events to LRF correlations}
\label{sm:sec:toy}

As a simple toy model linking a localization event to correlations with a quantum LRF, consider a delocalized single-particle system ${\cal S}$ on a spatial slice $\Sigma$, coarse-grained into sufficiently small, disjoint cells $B_i$ centered at $x_i$. In a regulated non-relativistic quantum-field description of ${\cal S}$, retaining one effective mode per cell, the reduced state of the retained mode in $B_i$ is
\begin{equation}
\rho_{B_i}\simeq p_i\,\rho_i^{(1)}+(1-p_i)\,\rho_i^{(0)}.
\label{sm:toy_local_state}
\end{equation}
Here $\rho_i^{(0)}:=|0\rangle_{B_i}\langle 0|$ and $\rho_i^{(1)}:=|1\rangle_{B_i}\langle 1|$ are the coarse-grained no-occupation and occupation states in $B_i$, respectively, and $p_i$ is the coarse-grained probability that the particle lies in $B_i$. The mixture in Eq.~(\ref{sm:toy_local_state}) results from tracing over the remaining cells and does not imply loss of coherence of the global one-particle state.

Now place in each cell a localized binary detector supported in $B_i$ that tests whether the excitation is present there. We take this detector to function as an LRF for this binary localization question. An idealized local measurement correlates the occupation variable with two perfectly distinguishable pointer records, $\sigma_i^{(0)}$ (no-click) and $\sigma_i^{(1)}$ (click), assumed to have orthogonal supports, $\sigma_i^{(0)}\sigma_i^{(1)}=0$. The corresponding post-interaction joint state is
\begin{equation}
\omega_{B_i}
=p_i\,\rho_i^{(1)}\otimes\sigma_i^{(1)}
+(1-p_i)\,\rho_i^{(0)}\otimes\sigma_i^{(0)},
\label{sm:toy_joint_state}
\end{equation}
while the reduced LRF state is
\begin{equation}
\sigma_{B_i}
=p_i\,\sigma_i^{(1)}+(1-p_i)\,\sigma_i^{(0)}.
\label{sm:toy_LRF_state}
\end{equation}
Equation~(\ref{sm:toy_joint_state}) makes explicit how a localization event is associated with a system--LRF correlation within a sufficiently small region. A related concrete realization of a quantum material reference frame is given in Ref.~\cite{sm:giacominiQR}, where a ``quantum ruler'' composed of harmonically interacting dipoles serves as a reference for position measurements. As noted in Ref.~\cite{sm:giacominiQR}, such a construction may be extended toward QFT by taking the continuum limit of the inter-site spacing.

In the spirit of relational QM~\cite{sm:Rovint}, each detector therefore functions as an LRF for the binary question ``is the particle in $B_i$?'', while the collection of these localized LRFs provides a coarse-grained ERF relative to which the particle's position becomes operationally well defined. 

\section{ Modular-correlation inequalities and the fixed-reference condition}
\label{sm:sec:modular}

We first derive, independently of Eq.~(\ref{constraint}), a modular-correlation inequality for the QFT setting considered in the main text and show that Eq.~(\ref{constraint}) is compatible with this bound but does not generically saturate it. We then show that its thermal specialization recovers the energy--correlation bound of Ref.~\cite{sm:Bruschi2015}. The exact identities in Eq.~(\ref{firstG}) give
\begin{align}
\delta I(\rho_B:\sigma_B)
&=
\delta S(\rho_B)+\delta S(\sigma_B)-\delta S(\omega_B)
\nonumber\\
&=
\delta\langle K\rangle
+\delta\langle K^{\scriptscriptstyle\rm LRF}\rangle
-S(\rho_B\Vert\rho_B^0)
-S(\sigma_B\Vert\sigma_B^0)
-\delta S(\omega_B).
\label{sm:generalmodularidentity}
\end{align}
Positivity of relative entropy therefore implies
\begin{equation}
\delta I(\rho_B:\sigma_B)+\delta S(\omega_B)
\leq
\delta\langle K\rangle
+\delta\langle K^{\scriptscriptstyle\rm LRF}\rangle.
\label{sm:generalmodularinequality}
\end{equation}
Since $\delta S(\omega_B)$ has no definite sign for an unrestricted joint state, Eq.~(\ref{sm:generalmodularinequality}) does not by itself yield an upper bound on $\delta I(\rho_B:\sigma_B)$ solely in terms of modular-energy variations.

A global inequality involving only mutual-information changes and modular-energy expectation values follows by considering a family of auxiliary global unitary operators acting on the isolated total system ${\cal S}+{\rm ERF}$. Consider joint states of the form
\begin{equation}
\omega[\rho]
= U[\rho]\,\omega^0\,U^\dagger[\rho],
\label{sm:physicaltotalstate}
\end{equation}
with ${\rm Tr}_{\rm ERF}\,\omega[\rho]=\rho$, so that $\rho$ is the reduced state of ${\cal S}$. Here $\omega^0$ is a fixed reference state, while $U[\rho]$ is a nonunique global-unitary parametrization not assigned any particular microscopic dynamics. This auxiliary class is introduced only to derive the correlation balance below and is not an additional assumption entering the derivation of the EE. 
Note that Eq.~(\ref{sm:physicaltotalstate}) refers to a global unitary acting on the isolated total system ${\cal S}+{\rm ERF}$ and should not be confused with the local parametrization $\omega_B[\rho_B]=U_B[\rho_B]\,\omega_B^0\,U_B^\dagger[\rho_B]$ introduced in the main text for the open subsystem $B$.

Defining $\omega_B={\rm Tr}_{\bar B}\omega$ and $\omega_{\bar B}={\rm Tr}_{B}\omega$, we have
\begin{equation}
I(\omega_B:\omega_{\bar B})
=
S(\omega_B)+S(\omega_{\bar B})-S(\omega).
\end{equation}
Since unitarity in Eq.~(\ref{sm:physicaltotalstate}) implies $\delta S(\omega)=0$, it follows that
$\delta I(\omega_B:\omega_{\bar B})
=
\delta S(\omega_B)+\delta S(\omega_{\bar B})$,
or equivalently
\begin{equation}
\delta S(\omega_B)
=
\delta I(\omega_B:\omega_{\bar B})
-\delta S(\omega_{\bar B}).
\end{equation}
Substitution into Eq.~(\ref{sm:generalmodularidentity}) gives
\begin{align}
\delta I(\rho_B:\sigma_B)
+\delta I(\omega_B:\omega_{\bar B})
={}&
\delta\langle K\rangle
+\delta\langle K^{\scriptscriptstyle\rm LRF}\rangle
+\delta S(\omega_{\bar B})
-S(\rho_B\Vert\rho_B^0)
-S(\sigma_B\Vert\sigma_B^0).
\label{sm:globalintermediate}
\end{align}

Defining the modular Hamiltonian of the exterior reference state by $K_{\bar B}=-\log\omega_{\bar B}^0$, its exact entropy identity reads 
\begin{equation}
\delta S(\omega_{\bar B})
=
\delta\langle K_{\bar B}\rangle
-S(\omega_{\bar B}\Vert\omega_{\bar B}^0).
\end{equation}
Substituting this equation into Eq.~(\ref{sm:globalintermediate}) gives the exact global correlation balance
\begin{align}
\delta I(\rho_B:\sigma_B)
+\delta I(\omega_B:\omega_{\bar B})
=
\delta\langle K\rangle
+\delta\langle K^{\scriptscriptstyle\rm LRF}\rangle
+\delta\langle K_{\bar B}\rangle
-S(\rho_B\Vert\rho_B^0)
-S(\sigma_B\Vert\sigma_B^0)
-S(\omega_{\bar B}\Vert\omega_{\bar B}^0).
\label{sm:globalmodularbalance}
\end{align}
Positivity of the three relative entropies then yields
\begin{equation}
\delta I(\rho_B:\sigma_B)
+\delta I(\omega_B:\omega_{\bar B})
\leq
\delta\langle K\rangle
+\delta\langle K^{\scriptscriptstyle\rm LRF}\rangle
+\delta\langle K_{\bar B}\rangle,
\label{sm:globalmodularinequality}
\end{equation}
which provides the corresponding bound for the general QFT setting.

Equation~(\ref{sm:globalmodularinequality}) is derived independently of Eq.~(\ref{constraint}). Note that, in the small-ball and ideal-LRF regime, Eq.~(\ref{constraint}) follows from Eq.~(\ref{sm:generalmodularidentity}) upon imposing the fixed-reference condition $\delta F=0$. Hence, within the auxiliary global-unitary construction of Eq.~(\ref{sm:physicaltotalstate}), states satisfying $\delta F=0$ [or, equivalently in this regime, Eq.~(\ref{constraint})] form a subset of the states that obey Eq.~(\ref{sm:globalmodularinequality}). Saturation of Eq.~(\ref{sm:globalmodularinequality}), however, requires additional restrictions not implied by Eq.~(\ref{constraint}); the bound is therefore generically strict. A microscopic realization selecting joint states that satisfy the fixed-reference condition lies beyond the scope of the present work.

The connection with Ref.~\cite{sm:Bruschi2015} becomes explicit in a particular thermal specialization of Eq.~(\ref{sm:globalmodularinequality}). The setup of Ref.~\cite{sm:Bruschi2015} consists of two initially uncorrelated thermal subsystems, $S_1$ and $S_2$, together with an initially uncorrelated thermal bath. We identify $S_1\leftrightarrow{\cal S}_B$, $S_2\leftrightarrow{\rm LRF}$, and let $\bar B$ play the role of the bath. If the three reference reduced states are thermal at the same inverse temperature $\beta$, their modular Hamiltonians take the thermal form
\begin{equation}
\delta\langle K\rangle=\beta\Delta E_{S_1},~~
\delta\langle K^{\scriptscriptstyle\rm LRF}\rangle=\beta\Delta E_{S_2},~~
\delta\langle K_{\bar B}\rangle=\beta\Delta E_{\rm bath}.
\end{equation}

Since the reference state is initially uncorrelated among the three subsystems~\cite{sm:Bruschi2015},
\begin{align}
\delta I(\rho_B:\sigma_B)=I(S_1:S_2),
\quad
\delta I(\omega_B:\omega_{\bar B})
=
I(S_1S_2:{\rm bath}).
\end{align}
Equation~(\ref{sm:globalmodularinequality}) therefore becomes
\begin{equation}
I(S_1:S_2)+I(S_1S_2:{\rm bath})
\leq
\beta\Delta E_{\rm total},
\label{sm:Bruschiglobal}
\end{equation}
where $\Delta E_{\rm total}=\Delta E_{S_1}+\Delta E_{S_2}+\Delta E_{\rm bath}$ is the work cost $W$ of Ref.~\cite{sm:Bruschi2015}. Equation~(\ref{sm:Bruschiglobal}) retains, in addition, the correlations generated between $S_1S_2$ and the bath. Since this contribution is nonnegative, dropping it gives
\begin{equation}
I(S_1:S_2)
\leq
\beta\Delta E_{\rm total},
\label{sm:deltaE}
\end{equation}
which is precisely the energy--correlation bound of Ref.~\cite{sm:Bruschi2015}. Thus, their thermal result follows as a particular limit of Eq.~(\ref{sm:globalmodularinequality}). In the general QFT setting considered here, however, the reference state may already contain correlations across $\partial B$, so $\delta I(\omega_B:\omega_{\bar B})$ need not be positive and cannot in general be discarded.

\section{ A possible relational UV regularization of the conditional entropy}
\label{sm:sec:uv}

Here we explore a further possibility allowed by the GIEH: the same ${\cal S}_B$--LRF correlations geometrically encoded by the metric may provide the UV structure required for the conditional entropy to admit a finite limit when a short-distance regulator $\epsilon$ is taken to zero. It is worth noting that this limit is not taken in the main text and is not required for the derivation of the EE. The regulated counterpart of the quantity entering the GIEH in Eq.~(\ref{GIEH}) is
\begin{equation}
S^{(\epsilon)}(\rho_B|\sigma_B)
=
S^{(\epsilon)}(\rho_B)
-
I^{(\epsilon)}(\rho_B:\sigma_B).
\label{sm:Cconditional}
\end{equation}
Although $S^{(\epsilon)}(\rho_B)$ and $I^{(\epsilon)}(\rho_B:\sigma_B)$ are finite for any fixed $\epsilon>0$, they may diverge separately as $\epsilon\to0$. Equation~(\ref{sm:Cconditional}) raises the possibility that, if the ${\cal S}_B$--LRF correlations extend to UV scales, their divergent parts cancel so that their difference admits a finite limit. 

We separate the UV contributions from the finite remainders as
\begin{equation}
S^{(\epsilon)}(\rho_B)
=
S_{\rm UV}^{(\epsilon)}
+
S_{\rm finite}(\rho_B)
+
o(1),
\label{sm:Sreg}
\end{equation}
and
\begin{equation}
I^{(\epsilon)}(\rho_B:\sigma_B)
=
I_{\rm UV}^{(\epsilon)}
+
I_{\rm finite}(\rho_B:\sigma_B)
+
o(1),
\label{sm:Ireg}
\end{equation}
as $\epsilon\to0$, where $S_{\rm finite}$ and $I_{\rm finite}$ denote the corresponding finite remainders and $o(1)$ represents terms vanishing in this limit. The regulated conditional entropy is therefore
\begin{align}
S^{(\epsilon)}(\rho_B|\sigma_B)
&=
S^{(\epsilon)}(\rho_B)
-
I^{(\epsilon)}(\rho_B:\sigma_B)
\nonumber\\
&=
S_{\rm UV}^{(\epsilon)}
-
I_{\rm UV}^{(\epsilon)}
+
S_{\rm finite}(\rho_B)
-
I_{\rm finite}(\rho_B:\sigma_B)
+
o(1).
\label{sm:Csplit}
\end{align}

We now ask whether the ${\cal S}_B$--LRF UV correlation structure can make the difference
$S_{\rm UV}^{(\epsilon)}-I_{\rm UV}^{(\epsilon)}$
approach a finite limit as $\epsilon\to0$. Finiteness alone would allow an arbitrary finite residue; we consider the more specific possibility
\begin{equation}
\lim_{\epsilon\to0}
\left[
S_{\rm UV}^{(\epsilon)}
-
I_{\rm UV}^{(\epsilon)}
\right]
=
\eta A ,
\label{sm:UVbalance}
\end{equation}
where $A$ denotes the area of $\partial B$ in the reference MSS. This condition requires the UV mutual information to reproduce the divergent part of the ${\cal S}_B$ entropy while leaving the finite residue $\eta A$. Thus, although $S_{\rm UV}^{(\epsilon)}$ and $I_{\rm UV}^{(\epsilon)}$ may diverge separately, their difference remains finite.

Taking the regulator to zero in Eq.~(\ref{sm:Csplit}) gives
\begin{align}
S(\rho_B|\sigma_B)
&\equiv
\lim_{\epsilon\to0}
S^{(\epsilon)}(\rho_B|\sigma_B)
\nonumber\\
&=
\eta A
+
S_{\rm finite}(\rho_B)
-
I_{\rm finite}(\rho_B:\sigma_B),
\label{sm:condfinite}
\end{align}
which is finite. After the identification $\eta=1/(4\hbar G)$ obtained in the main text, the combination $\eta A+S_{\rm finite}(\rho_B)$ in Eq.~(\ref{sm:condfinite}) has the same schematic area-plus-finite-entropy structure as the generalized entropy, although this observation does not constitute a microscopic derivation of it.

The GIEH in Eq.~(\ref{GIEH}) can now be expressed entirely in terms of finite quantities. The two sides, $S_g(\rho_B)$ and $S(\rho_B|\sigma_B)$, represent the same finite relational content in different descriptions. On the right-hand side, as given by Eq.~(\ref{sm:condfinite}), the UV balance leaves the finite residue $\eta A$, while the remaining explicit dependence on the LRF is carried by $I_{\rm finite}(\rho_B:\sigma_B)$. On the left-hand side, by contrast, the physical spacetime geometry encodes the relational information carried by the full regulated ${\cal S}_B$--LRF correlation structure, including both its UV and finite components, in accordance with the GIEH. Accordingly, we take the natural geometric counterpart of the finite UV residue $\eta A$ to be the corresponding area term $\eta A_g$, where $A_g$ denotes the boundary area in the deformed spacetime geometry. The difference $A_g-A$ then geometrically encodes the remaining finite correlation contribution quantified by $I_{\rm finite}(\rho_B:\sigma_B)$. We therefore write
\begin{equation}
S_g(\rho_B)
\approx
\eta A_g+
S_{\rm finite}(\rho_B).
\label{sm:Sgfinite}
\end{equation}

Combining Eqs.~(\ref{GIEH}),~(\ref{sm:condfinite}), and~(\ref{sm:Sgfinite}) then gives
\begin{equation}
\eta A_g
+
S_{\rm finite}(\rho_B)
\approx
\eta A
+
S_{\rm finite}(\rho_B)
-
I_{\rm finite}(\rho_B:\sigma_B),
\end{equation}
and hence
\begin{equation}
\eta\,\delta_g A
\approx
-I_{\rm finite}(\rho_B:\sigma_B),
\label{sm:GIEHfiniteArea}
\end{equation}
where $\delta_gA=A_g-A$. Here, $I_{\rm finite}$ is identified with the mutual information retained explicitly in the regulated effective description entering Eq.~(\ref{GIEH3}), including its background and perturbation-dependent contributions. Thus, Eq.~(\ref{sm:UVbalance}) provides a possible relational UV mechanism by which the conditional entropy admits a finite limit as the regulator is removed, while the GIEH geometrically encodes the explicit role of the LRF. In this sense, the area contribution underlying $\delta_g S(\rho_B)=\eta\,\delta_g A$ need not be introduced as an independent input from otherwise unspecified UV physics, but may instead arise from the ${\cal S}_B$--LRF correlation structure together with its geometric encoding by the GIEH. A microscopic realization of such a UV correlation structure is left for future work.

% Give the second bibliography its own width label and suppress repeated author notes.
\begingroup
\makeatletter
\let\@FMN@list\@empty
\patchcmd{\endthebibliography}{\label{LastBibItem}}{\label{LastSupplementalBibItem}}{}{}
\def\NAT@bibsetnum#1{\setlength{\topsep}{\z@}\NATx@bibsetnum{S3}}
\makeatother

\endgroup

\end{document}